\documentclass[twocolumn,prl]{revtex4}
\usepackage{bm}
\usepackage{epsfig}
\usepackage{amsmath}
\begin{document}
\title{Curve Crossing Problems: Analytically solvable model}
\author{Aniruddha Chakraborty \\
%EndAName
Department of Inorganic and Physical Chemistry, Indian Institute of Science,
Bangalore, 560012, India}
\date{\today }
\begin{abstract}
\noindent We propose an exactly solvable model for the two state
curve crossing problems. Our model assumes the coupling to be a delta
function. It is used to calculate the effect of curve crossing on
electronic absorption spectrum and resonance Raman excitation
profile.
\end{abstract}
\maketitle
\section{Introduction}
\noindent Nonadiabatic transition due to potential curve (or
surface) crossing is one of the most important mechanisms to
effectively induce electronic transitions in collisions
\cite{Naka1}. This is a very interdisciplinary concept and appears
in various fields of physics and chemistry, even in biology
\cite{Naka2}. The most typical examples are, of course, a variety
of atomic and molecular processes such as atomic and molecular
collisions, chemical reactions and molecular spectroscopic
processes. Attempts have been made to classify organic reactions
by curve crossing diagrams \cite{Shaik}. Other examples are
nuclear collisions and reactions in nuclear physics
\cite{Imanishi,Thiel}, dynamic processes on solid surfaces, energy
relaxation and phase transitions in condensed phase physics
\cite{Yoshimori,Engleman}, and electron and proton transfer
processes in biology \cite{Mataga,Dev}. Two state curve crossing
can be classified into the following two cases according to the
crossing scheme: (1) Landau-Zener (L.Z) case, in which the two
diabatic potential curves have the same signs for the slopes and
(2) non-adiabatic tunnelling (N.T) case, in which the diabatic
curves have the opposite sign for slopes. There is also a
non-crossing non-adiabatic transition called the
Rosen-Zener-Demkov type \cite{Naka1,Naka2}, in which two adiabatic
potentials are in near resonance at large $R$. The theory of
non-adiabatic transitions dates back to $1932$, when the
pioneering works for curve-crossing and non-crossing were
published by Landau \cite{Landau}, Zener \cite{Zener} and
Stueckelberg \cite{Stueckelberg} and by Rosen and Zener
\cite{Rosen} respectively. Since then numerous papers by many
authors have been devoted to these subjects, especially to curve
crossing problems. Two categories might be classified for finding
an exact analytical solution of the curve crossing problem. The
first is that an exact analytical solution can be obtained for the
whole region of the variable (say $x$ here, see in the next
section). For example, Osherov and Voronin solved the case where
two diabatic potentials are constant with exponential coupling
\cite{Voronin}. C. Zhu solved the case where two diabatic
potentials are exponential with exponential coupling
\cite{Nikitinmodel}. The second is that an exact analytical
solution is not possible for the whole region of the variable, but
is possible for the asymptotic region. Then, physical quantities
such as eigenvalues, scattering matrices  can still be solved in
an exact analytical form, providing that the connection problem of
the asymptotic solution is known. The Stokes phenomenon
\cite{Stokes} of asymptotic solution of the ordinary differential
equation provides a powerful tool to deal with these kinds of
problems \cite{Heading, Hinton, Zhu}. Generalizing the real
variable to the complex variable and tracing the asymptotic
solution around the complex plane, the connection matrix which
connects the asymptotic solution in the complex plane can be
expressed in terms of Stokes constants. Recent work by Zhu and
Nakamura \cite{Zhu} found an exact analytical solution of the
Stokes constants for the second-order ordinary differential
equation with the coefficient function as the fourth-order
polynomial. In this way, exact analytical solutions of scattering
matrices were obtained for the two state linear curve crossing
problem with constant coupling \cite{Aquilanti}.
\section{The model}
\noindent We consider two diabatic curves, crossing each other.
There is a coupling between the two curves, which causes
transitions from one curve to another. This transition would occur
in the vicinity of the crossing point. In particular, it will
occur in a narrow range of $x$, given by
\begin{equation}
\label{1}\left|V_1(x)-V_2(x)\right|\simeq \left|V(x_c)\right|.
\end{equation}
where $x$ denotes the nuclear coordinate and $x_c$ is the crossing point. $%
V_1$ and $V_2$ are the diabatic curves and $V$ represent the
coupling between them. Therefore it is interesting to analyze a
model, where coupling is localized in space near $x_c$. Thus we
put
\begin{equation}
\label{2}V(x)=K_0\delta (x-x_c),
\end{equation}
here $K_0$ is a constant. This model has the advantage that it can
be exactly solved (delta function sink model for electronic
relaxation in solution is known in literature \cite{Sebastian}).
One can roughly estimate the value of $K_0$ as follows. If the two
curves are approximated as linear near the vicinity of crossing,
then the energy separation between the two is $\approx \left[
\frac \partial {\partial x}(V_1-V_2)\right] _{x_c}(x-x_c)$. This
will be of the order of $\Delta V(x_c)$, if $(x-x_c)$ takes the
value $\simeq \frac{\Delta V(x_c)}{\left( \frac \partial {\partial
x}(V_1-V_2)\right) _{x_c}}$. So it is reasonable to put $K_0=\left| \frac{%
2V(x_c)}{\left( \frac \partial {\partial x}(V_1-V_2)\right) _{x_c}}\right|
\times V(x_c).$
\section{Exact analytical solution}
\noindent We start with a particle moving on any of the two
diabatic curves. The problem is to calculate the probability that
the particle will still be on any one of the diabatic curves after
a time $t$. We write the probability amplitude as
\begin{equation}
\label{3}\Psi (x,t)=\left(
\begin{array}{c}
\psi _1(x,t) \\
\psi _2(x,t)
\end{array}
\right) ,
\end{equation}
where $\psi _1(x,t)$ and $\psi _2(x,t)$ are the probability amplitude for
the two states. $\Psi (x,t)$ obey the time dependent Schr$\stackrel{..}{o}$
dinger equation (we take $\hbar =1$ here and in subsequent calculations)
\begin{equation}
\label{4}i\frac{\partial \Psi (x,t)}{\partial t}=H\Psi (x,t).
\end{equation}
$H$ is defined by
\begin{equation}
\label{5}H=\left(
\begin{array}{cc}
H_1(x) & V(x) \\
V(x) & H_2(x)
\end{array}
\right) ,
\end{equation}
where $H_i(x)$ is
\begin{equation}
\label{6}H_i(x)=-\frac 1{2m}\frac{\partial ^2}{\partial x^2}+V_i(x).
\end{equation}
We find it convenient to define the half Fourier Transform $\overline{\Psi }%
(\omega )$ of $\Psi (t)$ by
\begin{equation}
\label{9}\overline{\Psi }(\omega )=\int_0^\infty \Psi (t)e^{i\omega t}dt.
\end{equation}
Half Fourier transformation of Eq. (\ref{4}) leads to
\begin{equation}
\label{10}\left(
\begin{array}{c}
\overline{\psi }_1(\omega ) \\ \overline{\psi }_2(\omega )
\end{array}
\right) =i\left(
\begin{array}{cc}
\omega -H_1 & V \\
V & \omega -H_2
\end{array}
\right) ^{-1}\left(
\begin{array}{c}
\psi _1^{}(0) \\
\psi _2(0)
\end{array}
\right) .
\end{equation}
This may be written as
\begin{equation}
\label{11}\overline{\Psi }(\omega )=iG(\omega )\Psi (0).
\end{equation}
$G(\omega )$ is defined by $(\omega -H)$ $G(\omega )=I$. In the position
representation, the above equation may be written as
\begin{equation}
\label{12}\overline{\Psi }(x,\omega )=i\int_{-\infty }^\infty G(x,x_0;\omega
)\overline{\Psi }(x_0,\omega )dx_0,
\end{equation}
where $G(x,x_0;\omega )$ is
\begin{equation}
\label{13}G(x,x_0;\omega )=\langle x|(\omega -H)^{-1}|x_0\rangle .
\end{equation}
Writing
\begin{equation}
\label{14}G(x,x_0;\omega )=\left(
\begin{array}{cc}
G_{11}^{}(x,x_0;\omega ) & G_{12}^{}(x,x_0;\omega ) \\
G_{21}^{}(x,x_0;\omega ) & G_{22}^{}(x,x_0;\omega )
\end{array}
\right)
\begin{array}{cc}
&  \\
&
\end{array}
\end{equation}
and using the partitioning technique \cite{Lowdin} we can write
\begin{equation}
\label{15}G_{11}^{}(x,x_0;\omega )=\langle x|[\omega -H_1-V(\omega
-H_2)^{-1}V]^{-1}|x_0\rangle.
\end{equation}

The above equation is true for any general $V$. This expression
simplify considerably if $V$ is a delta function located at $x_c$.
In that case $V$ may be written as $V=$ $K_0S=K_0$ $|x_c\rangle $
$\langle x_c|$. Then
\begin{equation}
\label{16}G_{11}^{}(x,x_0;\omega )=\langle x|[\omega
-H_1-K_0^2G_2^0(x_c,x_c;\omega )S]^{-1}|x_0\rangle ,
\end{equation}
where
\begin{equation}
\label{17}G_2^0(x,x_0;\omega )=\langle x|(\omega -H_2^{})^{-1}|x_0\rangle ,
\end{equation}
and corresponds to propagation of the particle starting at $x_0$
on the second diabatic curve, in the absence of coupling to the
first diabatic curve. Now we use the operator identity
\begin{widetext}
\begin{equation}
\label{18} (\omega -H_1-K_0^2G_2^0(x_c,x_c;\omega )S)^{-1}=
(\omega -H_1)^{-1}+(\omega -H_1)^{-1}K_0^2G_2^0(x_c,x_c;\omega
)S[\omega -H_1-K_0^2G_2^0(x_c,x_c;\omega )S]^{-1}.
\end{equation}
\end{widetext}
Inserting the resolution of identity $I=\int_{-\infty }^\infty dy$
$ |y\rangle $ $\langle y|$ in the second term of the above
equation, we arrive at
\begin{align}
\label{20}G_{11}^{}(x,x_0;\omega)=&G_1^0(x,x_0;\omega)+K_0^2G_1^0(x,x_c;\omega)\\
& \times G_2^0(x_c,x_c;\omega)G_{11}(x_c,x_0;\omega).
\end{align}
\newline
Putting $x=x_c$ in Eq. (\ref{20}) and solving for
$G_{11}(x_c,x_0;\omega )$ gives
\begin{align}
\label{21}G_{11}(x_c,x_0;\omega )=&G_1^0(x_c,x_0;\omega )\\&
\times [1-K_0^2G_1^0(x_c,x_c;\omega )G_2^0(x_c,x_c;\omega )]^{-1}.
\end{align}
\noindent This when substituted back into Eq. (\ref{20}) gives
\begin{widetext}
\begin{equation}
\label{22} G_{11}(x,x_0;\omega )=
G_1^0(x,x_0;\omega)+\frac{K_0^2G_1^0(x,x_c;\omega)G_2^0(x_c,x_c;\omega
)G_1^0(x_c,x_0;\omega )}{ 1-K_0^2G_1^0(x_c,x_c;\omega
)G_2^0(x_c,x_c;\omega )}.
\end{equation}
\end{widetext}
Using the same procedure one can get
\begin{equation}
\label{23}
\begin{array}{c}
G_{12}^{}(x,x_0;\omega )=\frac{K_0G_1^0(x,x_c;\omega )G_2^0(x_c,x_0;\omega )%
}{1-K_0^2G_1^0(x_c,x_c;\omega )G_2^0(x_c,x_c;\omega )}.
\end{array}
\end{equation}
Similarly one can derive expressions for $G_{22}^{}(x,x_0;\omega )$ and $%
G_{21}^{}(x,x_0;\omega )$. Using these expressions for the Green's
function in Eq. (\ref{11}) we can calculate $\overline{\Psi
}(\omega )$ explicitly.
\newline
The expressions that we have obtained for $\overline{\Psi }(\omega
)$ are quite general and are valid for any $V_1(x)$ and $V_2(x)$.
However, their utility is limited by the fact that one must know
$G_1^0(x,x_0;\omega )$ and $G_2^0(x,x_0;\omega )$. It is possible
to find $G_i^0(x,x_0;\omega )$ only in a few limited cases and the
harmonic oscillator is one of them.
\section{Electronic Absorption Spectra and Resonance Raman Excitation
Profile} In this section we apply the method to the problem
involving harmonic potentials. We consider a system of three
potential energy curves, ground electronic state and two
`crossing' excited electronic states (electronic transition to one
of them is assumed to be dipole forbidden and while it is allowed
to the other) \cite{Zink,ZinkPRL}. We calculate the effect of
`crossing' on electronic absorption spectra and on resonance Raman
excitation profile. The propagating wave functions on the excited
state potential energy curves are given by solution of the time
dependent Schr\"{o}dinger equation
\begin{figure}[h]
\centering\epsfig{file=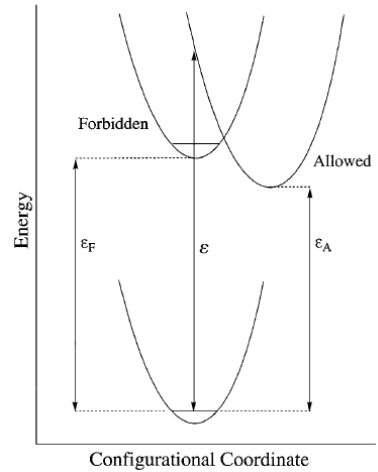,width=0.6\linewidth}
\caption{Schematic diabatic potential energy curves that
illustrate the model. The forbidden state is labeled "F"; the
allowed state is labeled "A".} \label{Curveapply}
\end{figure}
\begin{widetext}
\begin{equation}
\label{N33}i\frac \partial {\partial t}\left(
\begin{array}{c}
\psi _1^{vib}(x,t) \\
\psi _2^{vib}(x,t)
\end{array}
\right)=\left(
\begin{array}{cc}
H_{vib,e1}(x) & V_{12}(x) \\
V_{21}(x) & H_{vib,e2}(x)
\end{array}
\right) \left(
\begin{array}{c}
\psi _1^{vib}(x,t) \\
\psi _2^{vib}(x,t)
\end{array}
\right).
\end{equation}
\end{widetext}
In the above equation $H_{vib,e1}(x)$ and $H_{vib,e2}(x)$
describes the vibrational motion of the system in the first
electronic excited state (allowed) and second electronic excited
state (forbidden) respectively
\begin{equation}
\label{N34}H_{vib,e1}(x)=-\frac 1{2m}\frac{\partial ^2}{\partial
x^2}+\frac 12m\omega _A^2(x-a)^2
\end{equation}
and
\begin{equation}
\label{N35}H_{vib,e2}(x)=-\frac 1{2m}\frac{\partial ^2}{\partial
x^2}+\frac 12m\omega _F^2(x-b)^2.
\end{equation}
In the above $m$ is the oscillator's mass, $\omega _A$ and $\omega
_F$ are the vibrational frequencies on the allowed and forbidden
states and $x$ is the vibrational coordinate. Shifts of the
vibrational coordinate minimum upon excitation are given by $a$
and $b$, and $V_{12}$ ($V_{21}$) represent coupling between the
two harmonic potentials which is taken to be
\begin{equation}
\label{N36}V_{21}(x)=V_{12}(x)=K_0\delta (x-x_c),
\end{equation}
where $K_0$ represent the strength of the coupling.
\begin{figure}[h]
\centering\epsfig{file=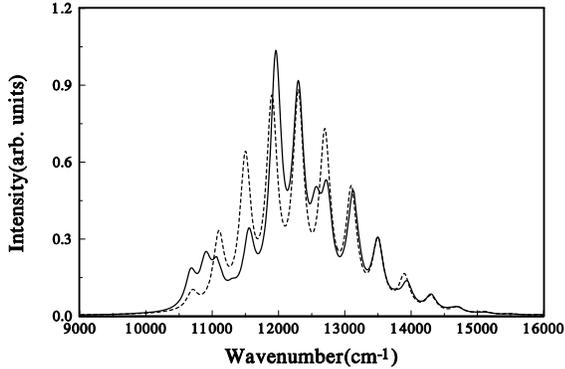,width=1\linewidth} \caption{
Calculated electronic absorption spectra with coupling (solid
line) and without coupling (dashed line). Here the values for the
simulations are $\omega _0=\omega _A=\omega _F=400\:cm^{-1}$,
$\Gamma =100\:cm^{-1}$, $\varepsilon _A=10700\:cm^{-1}$,
$\varepsilon _F=11000\:cm^{-1}$, $K_0=5.9643 \times
10^{-15}\:erg.\AA$, and $x_c=0.0895\:\AA$.} \label{Elec}
\end{figure}
The intensity of electronic absorption spectra is given by
\cite{Zink,Heller}
\begin{eqnarray}
\label{N41}I_A(\omega )\propto & Re[\int_{-\infty }^\infty
dx\int_{-\infty }^\infty dx_0\Psi_i ^{vib*^{}}(x)\nonumber
\\ & iG(x,x_0;\omega +i\Gamma )\Psi_i ^{vib}(x_0)],
\end{eqnarray}
where
\begin{equation}
\label{N42}G(x,x_0;\omega +i\Gamma )=\langle x|[(\omega_0/2+\omega
-\omega _{eg})+i\Gamma -H_{vib,e}]^{-1}|x_0\rangle .
\end{equation}
and
\begin{equation}
\label{N42a} H_{vib,e}=\left(
\begin{array}{cc}
H_{vib,e1}(x) & K_0 |x_c\rangle\langle x_c| \\
K_0  |x_c\rangle\langle x_c| & H_{vib,e2}(x)
\end{array}
\right)
\end{equation}
Here, $\Gamma$ is a phenomenological damping constant which
account for the life time effects. $\Psi_i^{vib}(x,0)$ is given by
\begin{equation}
\label{N42b}\Psi _i^{vib}(x,0)=\left(
\begin{array}{c}
\chi_i(x) \\
0
\end{array}
\right),
\end{equation}
where $\chi_i(x)$ is the ground vibrational state of the ground
electronic state, $\omega_0$ is the vibrational frequency on the
ground electronic state, $\varepsilon_A$ is the energy difference
between the excited (allowed) and ground electronic state, and for
the forbidden electronic state it's value is $\varepsilon_F$.
Similarly resonance Raman scattering intensity can be expressed in
terms of Green's function and is given by \cite{Heller,Zink}.
\begin{eqnarray}
\label{N53} I_R(\omega )\propto &|\int_{-\infty }^\infty
dx\int_{-\infty
}^\infty dx_0\Psi _f^{vib*}(x,0)\nonumber\\
& iG(x,x_0;\omega+i\Gamma)\Psi _i^{vib}(x_0,0)|^2.
\end{eqnarray}
In the above $\Psi _f^{vib}(x,0)$ is given by
\begin{equation}
\label{N53a}\Psi _f^{vib}(x,0)=\left(
\begin{array}{c}
\chi _f(x) \\
0
\end{array}
\right),
\end{equation}
where $\chi _f(x)$ is the final vibrational state of the ground
electronic state. As $G_i^0(x,x_0;\omega)$ for the harmonic
potential is known \cite{Grosche}, we can calculate
$G(x,x_0;\omega)$. We use Eq. (\ref{N53}) to calculate the effect
of curve crossing on resonance Raman excitation profile.
\begin{figure}[h]
\centering\epsfig{file=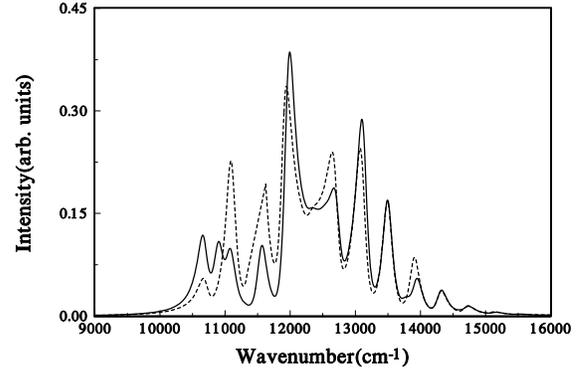,width=1\linewidth}
\caption{Calculated resonance Raman excitation profile for
excitation from the ground vibrational state to the first excited
vibrational state, with coupling (solid line) and without coupling
(dashed line). Here the values for the simulations are $\omega
_0=\omega _A=\omega _F=400\:cm^{-1}$, $\Gamma =100\:cm^{-1}$,
$\varepsilon _A=10700\:cm^{-1}$, $\varepsilon _F=11000\:cm^{-1}$,
$K_0=5.9643 \times 10^{-15}\:erg.\AA$, and $x_c=0.0895\:\AA$.}
\label{Raman}
\end{figure}

\subsection{Results using the model}
In the following we give results for the effect of curve crossing
on electronic absorption spectrum and resonance Raman excitation
profile in the case where one dipole allowed electronic state
crosses with a dipole forbidden electronic state as in Fig.
\ref{Curveapply}. As in \cite{Zink}, the ground state curve is
taken to be a harmonic potential energy curve with its minimum at
zero. The curve is constructed to be representative of the
potential energy along a metal-ligand stretching coordinate. We
take the mass as $35.4\:amu$ and the vibrational wavenumber as
$400\:cm^{-1}$ \cite{Zink} for the ground state. The first
diabatic excited state potential energy curve is displaced by
$0.197\:\AA$ and is taken to have a vibrational wavenumber of
$400\:cm^{-1}$. Transition to this state is allowed. The minimum
of the potential energy curve is taken to be above
$10700\:cm^{-1}$ of that of the ground state curve. The second
diabatic excited state potential energy curve is taken to be an
un-displaced excited state. On that potential energy curve, the
vibration is taken to have same wavenumber of $400\:cm^{-1}$. Its
minimum is $11000\:cm^{-1}$ above that of the ground state curve.
Transition to this state is assumed to be dipole forbidden. The
two diabatic curves cross at an energy of $11673.6\:cm^{-1}$ with
$x_c=0.0895\:\AA$. Value of $K_0$ we use in our calculation is
$K_0=5.9643 \times 10^{-15}\:erg.\AA$. The lifetime of both the
excited states are taken to be $100\:cm^{-1}$. The calculated
electronic absorption spectra is shown in Fig. \ref{Elec}. The
profile shown by the dashed line is in the absence of any coupling
to the second potential energy curve. The full line has the effect
of coupling in it. The calculated resonance Raman excitation
profile is shown in Fig. \ref{Raman}. The profile shown by the
full line is calculated for the coupled potential energy curves.
The profile shown by the dashed line is calculated for the
uncoupled potential energy curves. It is seen that curve crossing
effect can alter the absorption and Raman excitation profile
significantly. However it is the Raman excitation profile that is
more effected. Even though our calculations are for a simple model
(which has the convenience that everything can be found
analytically), similar conclusions are valid for more complicated
couplings \cite{Zink}.
\section{Conclusions}
\noindent We have proposed a simple model for the two state curve
crossing problem which can be analytically solved. Our solution is
quite general and is valid for any potentials for which Green's
functions for the motion in the absence of coupling is known. The
model is used to calculate the effect of curve crossing on
electronic absorption spectrum and on resonance Raman excitation
profile. We find that Raman excitation profile is affected much
more by the crossing, than the electronic absorption spectrum.
\section{acknowledgments}
\noindent The author thanks Prof. K. L. Sebastian for giving the
idea of solving this problem and also for valuable suggestions.

\end{document}